\documentclass[amsmath,amssymb,pre]{revtex4}

\usepackage{graphicx} 
\usepackage{float}
\usepackage{xcolor}
\usepackage{comment}
\usepackage{url}

\begin{document}
\title{From the perceptron to the cerebellum}
\author{Nicolas Brunel}
\email{nicolas.brunel@unibocconi.it}
\affiliation{Department of Computing Sciences, Universit\`a Bocconi, Milan, Italy}
\author{Vincent Hakim}
\email{vincent.hakim@ens.fr}
\affiliation{Laboratoire de Physique de l’École normale supérieure, ENS, Université PSL, CNRS, Sorbonne Université, Université Paris Cité, Paris, France}
\author{Jean-Pierre Nadal}
\email{jean-pierre.nadal@ens.fr}
\affiliation{Laboratoire de Physique de l’École normale supérieure, ENS, Université PSL, CNRS, Sorbonne Université, Université Paris Cité, Paris, France}
\affiliation{Centre d'Analyse et de Math\'ematique Sociales, EHESS, CNRS, Paris, France}

\date{\today}
\begin{abstract}
The perceptron has served as a prototypical neuronal learning machine in the physics community interested in neural networks and artificial intelligence, which included Gérard Toulouse as one of its prominent figures. It has also been used as a model of Purkinje cells of the cerebellum, a brain structure involved in motor learning, in the early influential theories of David Marr and James Albus. We review these theories, more recent developments in the field, and highlight questions of current interest.\\$\;$\\
This paper is a preprint of an article published as part of a thematic issue "Gérard Toulouse, a life of discovery and commitment" (\url{https://doi.org/10.5802/crphys.sp.1}) of the {\em Comptes Rendus Physique de l'Académie des Sciences} (revue of the French National Academy of Sciences). 
\end{abstract}

\maketitle

\section{Introduction}
It is a great pleasure to contribute to this volume honoring the memory and scientific contributions of Gérard Toulouse. It is also an opportunity to acknowledge our gratitude for his scientific leadership, which has played a key role in our own scientific paths. Gérard was an early leading figure in the physics of spin glasses which caught his interest soon after the pioneering work of Sherrington and Kirkpatrick \cite{sherrington75}.
As recalled in other articles in this volume, some of his  contributions include the analysis of the coexistence of ferromagnetic and spin-glass order \cite{gabay81} as well as the formal introduction of the concept of frustration \cite{toulouse77}. He strongly  pushed this line of research after his move from Orsay to create a statistical physics research group at Ecole Normale Sup\'erieure (ENS). The fruitful collaboration with the Rome group of Giorgio Parisi led to the discovery of ultrametricity in high dimensional spin glasses \cite{mezard86}. At about the same time, the introduction of the closely related attractor neural network model by John Hopfield \cite{hopfield82} drew the attention of the spin-glass community. Gérard Toulouse soon became a strong advocate of the importance of such models for biology and, in particular, for helping to realize his dream of understanding how the brain works. He seeked to collaborate with biologists and to attract younger researchers to this promising line of research~\cite{toulouse86,nadal86}. He started to investigate the link between neural coding and information theory \cite{nadal90}, using the perceptron as a prototypical learning machine \cite{brunel92}. These topics are now important components of computational neuroscience, which has developed into a full research field, starting from these early beginnings. 

In the present contribution, we would like to highlight work
that followed from G\'erard Toulouse's initial lead. We have chosen to focus on the cerebellum, as one of the brain structures in which perceptron theory has had an important impact.  Classic early theories make key use of concepts such as perceptron learning, as we recall below, and the cerebellum has since been the subject of much interdisciplinary research between biologists and physicists, at the ENS in particular.

\section{The cerebellar anatomy and Marr-Albus-Ito theory.}

The cerebellum is a brain structure that is located in the back of the brain, between the  cerebral cortex and the brain stem \cite{shepherd03}. As its name indicates, it is smaller than the cerebral cortex in terms of size, though it contains more neurons. It is classically thought  to be a primary site of motor learning and motor control \cite{raymond18}. It is specially involved  in the production of precise and smooth movements and in motor coordination. It has also been shown to be involved in purely cognitive tasks such as word associations, and is implicated in several neurological disorders \cite{schmahmann19}.

\begin{figure}
\begin{tabular}{ll}
{\bf A} & {\bf B} \\
\begin{minipage}{0.49\textwidth}
\includegraphics[width=\textwidth]{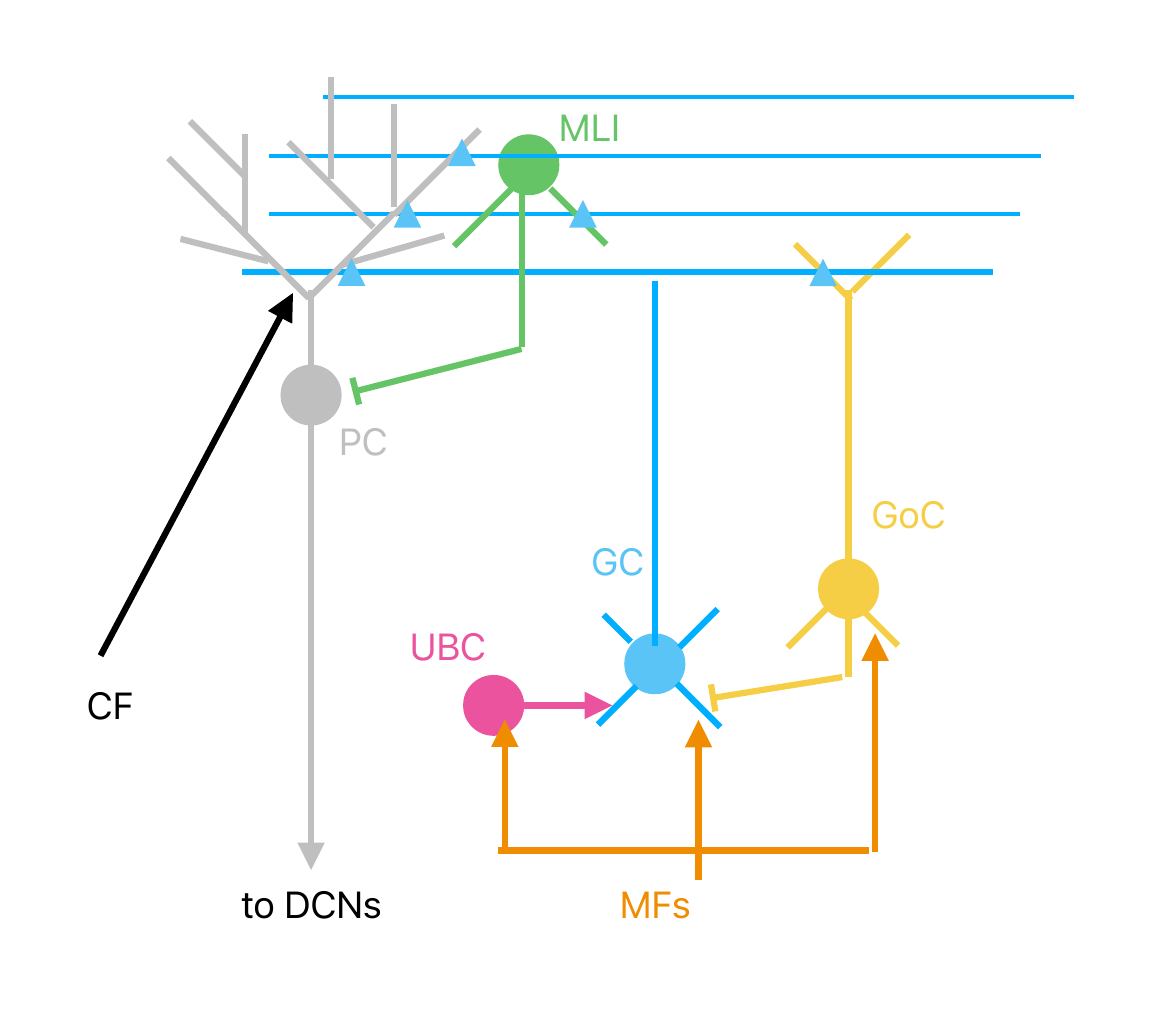} 
\end{minipage} & 
\begin{minipage}{0.49\textwidth}
\includegraphics[width=0.8\textwidth]{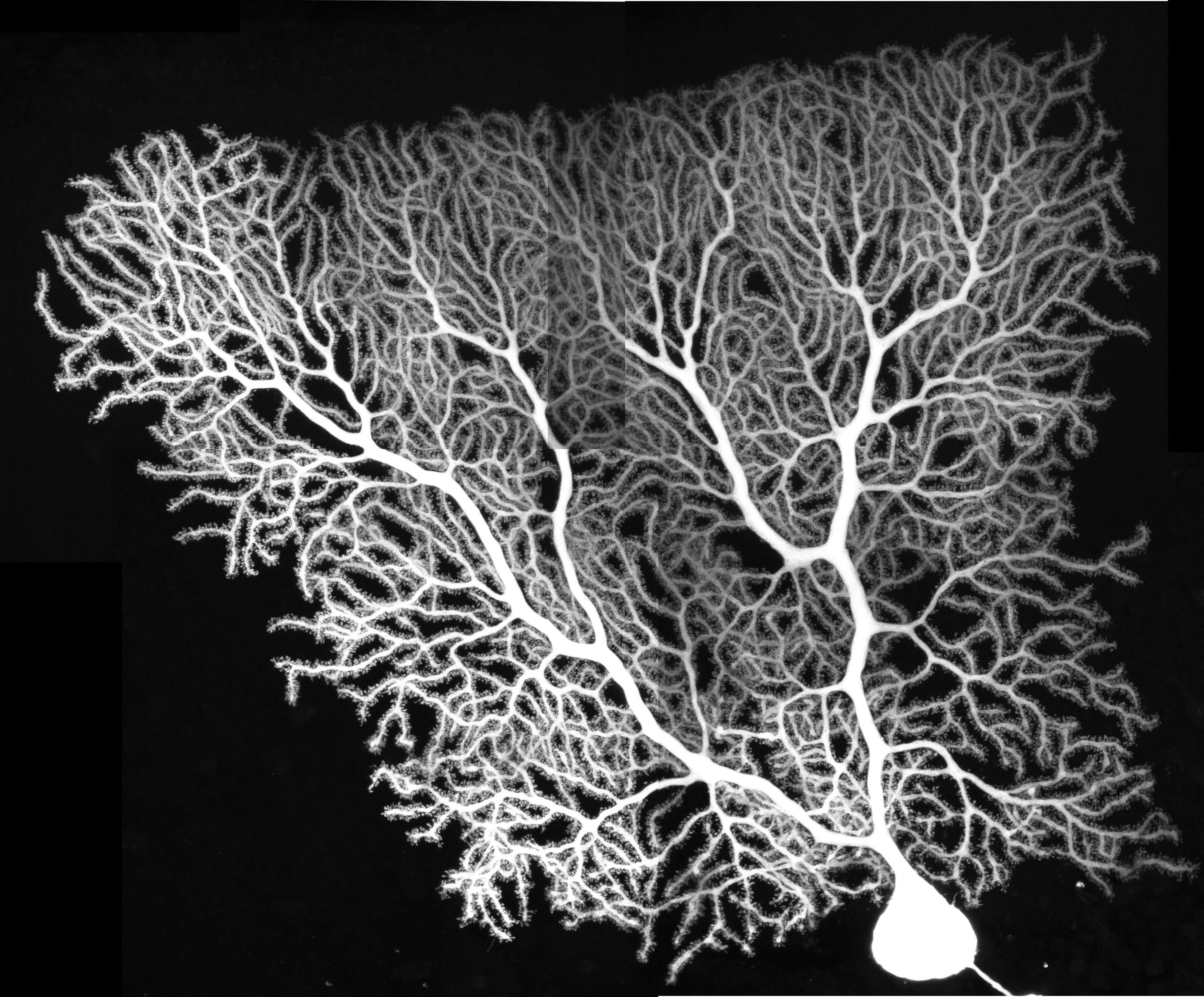}
\end{minipage}
\end{tabular}
\caption[]{Anatomy of the cerebellum. A: Cell types and connectivity showing Purkinje cells (PC), granule cells (GrC), unipolar brush cells (UBC), molecular layer interneurons (UBC)  and Golgi cells (GoC) as well as mossy fibers (MF) and climbing fibers (CF) coming from the inferior olive (from \cite{hull22}). B: Purkinje cell anatomy (courtesy of Boris Barbour).}
\label{fig:cerebellar-anatomy}
\end{figure}

The highly regular anatomy of the cerebellum  (Fig.~\ref{fig:cerebellar-anatomy}A) distinguishes it from other brain structures and has strongly contributed to attract the attention of experimentalists and theoreticians. The Purkinje cells are the central cells of the cerebellum and organize its structure \cite{shepherd03,hull22}. They are inhibitory neurons with a very large, extensive and unusually planar dendritic arbor, as shown in Fig.~\ref{fig:cerebellar-anatomy}B. They are organized in a regular array and constitute the only output of the 
 cerebellar cortex.
Their axons target deeper neural structures, collectively known as the deep cerebellar nuclei. Remarkably, there are only two kinds of 
 glutamatergic inputs \cite{shepherd03,hull22} to the cerebellar cortex\footnote{The cerebellum also receives monoaminergic \cite{anden67,ChineAugustine2025} and cholinergic \cite{Jaarsma97,Fore2020} projections.}.

The first type of inputs, provided by mossy fibers (MFs), convey sensory and proprioceptive information from all body parts. These inputs are relayed to the Purkinje cells (PCs) by the granule cells (GCs) which are excitatory neurons. GCs are so numerous that they account for about half of all the neurons in the brain. As shown in Fig.~\ref{fig:cerebellar-anatomy}A, the granule cell axons climb radially into the cerebellar cortex and bifurcate horizontally in the most superficial cerebellar layer (called the molecular layer). These so-called parallel fibers (PFs), traverse perpendicularly  a few hundred Purkinje cell dendritic arbors, making with each a synapse with about 50\% probability \cite{harvey88,harvey91}. This leads each Purkinje cell to receive inputs from an unusually large number of granule cells ($\sim 10^5$). Besides these two types of cells, the cerebellar cortex contains other types of neurons. Mossy fibers target unipolar brush cells (UBCs), excitatory cells that in turn target granule cells. Both mossy fibers and parallel fibers target Golgi cells, which themselves provide feedback inhibition to the granule cells. Parallel fibers also target other inhibitory interneurons (molecular layer interneurons, sometimes classified as basket and stellate cells) that inhibit Purkinje cells in a feedforward manner. Finally, rarer and less studied cell types \cite{hull22} (Lugaro cells, Candelabrum cells, and Globular cells) are also present in Purkinje cell layers, but their roles are still mysterious and are typically ignored in theoretical models.

The second excitatory input to the cerebellar cortex 
is provided by the axons (the ``climbing fibers'') of cells in a small structure called the inferior olive (IO). As shown in Fig.~\ref{fig:cerebellar-anatomy}, each Purkinje cell receives input from a single IO axon, while each IO cell innervates $\sim 7$ Purkinje cells  \cite{sugihara01}.

In humans, there are a total of about 5.$10^{10}$ granule cells,  15.$10^6$ Purkinje cells, 2.5 $10^8$ mossy fibers and $10^6$ IO cells \cite{shepherd03}.
Although it is known that different parts of the cerebellar cortex 
are associated with different modalities and the control of different body parts, its uniform cellular connectivity suggests that general principles may guide the design of its structure.

This led David Marr and James Albus  to propose  their classic theories, soon after anatomical data was obtained and synthesized \cite{eccles67}.
Marr \cite{marr69} first proposed that granule cells serve to recode mossy fiber inputs and expand them in a higher dimensional space to facilitate the discrimination of closely similar MF inputs. He further proposed that Golgi cells serve to maintain a sparse activity of granule cells for the same purpose. Crucially, he interpreted Purkinje cells as
perceptrons, with their activity being set by comparing the linear sum of the inputs they receive from granule cells to a threshold controlled by the stellate and basket cells.

Albus' account of the cerebellar structure is similar with an important difference \cite{albus71}. Marr suggested that the cerebellar refinement of movement was due to a change in the origin of Purkinje cell discharge. Namely, he proposed that heterogeneous plasticity would allow the  progressive replacement of  Purkinje cells discharge from complex spikes (CSs), driven via the IO, to a more precisely modulated discharge by simple spikes, refined with the help of granule cell discriminating power. This led him to propose that joint CS-PF inputs in a Purkinje cell would lead to a {\em potentiation} of the active PF synapses. Although this view has some appeal, Albus criticized it on account of stability, pointing out that PF-PC potentiation would be at risk of being a runaway process. Instead, he proposed that the single climbing fiber innervating a Purkinje cell 
would act as a
"teacher", namely signaling PC firing errors, and that the joint activation of CS-PF inputs should lead to {\em depression} of PC-PF synapses.
Electrophysiological experiments in {\em in vitro} preparations, starting from the pioneering work of Ito's group \cite{ito82}, have substantiated Albus's views. Further support has been provided by recordings {\em in vivo} during simple tasks dependent on cerebellar learning \cite{yang14}.

One prototypical such task is the eyeblink conditioning experiment \cite{thompson98}. An air puff on the eye provokes eyelid closure as a defensive reflex. This unconditioned stimulus can be paired in a Pavlovian fashion with a neutral anticipatory stimulus such as the turning on of a light or of a sound. After multiple repetitions of this pairing, eyelid closure appears after the conditioned stimulus (the neutral stimulus) 
in anticipation of the air puff (similarly as dog salivation upon bell ringing in anticipation of food in Pavlov's famous experiment). Numerous experiments, notably from Thomson's lab \cite{thompson98}, have shown that this conditioning is dependent on the cerebellum. For instance, blocking the output of the cerebellum prevents the anticipatory response expression. The neutral stimulus can also be replaced by direct stimulation of parallel fibers.
Recordings from Purkinje cells in a specific region of the cerebellum show that initially a burst of complex spikes is provoked by the air puff. Upon conditioning, these complex spikes disappear, which is interpreted as a suppression of the error of non-closure of the eyelid.  

Marr-Albus-Ito theory now usually refers to the view of the cerebellar cortex  
as learning in a supervised fashion from a teacher provided by inferior olive cells, helped by the associated granule cells representational expansion of inputs and depression at the PF-PC synapse 
upon joint stimulation by CF and granule cells. Since these original proposals, the theory has been examined and expanded in multiple directions, following progress in experimental and computational techniques. We describe some of this work in the following sections.\\

\section{Granule cells and expansion in high dimensions.}
One useful geometric viewpoint, which has become very popular in recent years \cite{churchland12}, consists in representing the activity of a set of  $N$ neurons as a point in a multidimensional space, with each axis representing
the activity of a neuron. If rate units are considered, the activity point  will represent $(r_1,\cdots, r_N)$  where $r_j$s are the firing rates of different neurons. If spikes in small discrete time bins are considered, the activity of the set at each time bin can be described as $(\sigma_1,\cdots,\sigma_N)$, with
 $\sigma_i=1$ or $0$ depending on whether neuron $i$ emits a spike or not in the time bin considered. The activity is then restricted to the vertices of a hypercube in $N$ dimensions. When the activities of the different units are correlated, the activity of the network typically takes place in a subspace of smaller dimension than $N$, as has been observed in recent multi-neuron recordings \cite{gao15,gallego17,chung21}. 
 
 Marr and Albus both made an early use of this geometric viewpoint to note that the much higher number of granule cells as compared to mossy fibers means that the mossy fiber activity space is projected in the much higher dimensional activity space of the granule cells. They proposed the influential idea that the benefit of this expansion, and its underlying reason, is that different activity patterns are easier to discriminate if they occur in a space of larger dimension.

Marr's simple reasoning 
goes as follows. Consider two sets of (mossy fiber) patterns $\mathbf{p}_1$ and $\mathbf{p}_2$ with 0 (inactive) or 1 (active) entries and a coding level, the probability that an entry is active, such that there are $L$ active entries in each pattern.  The cosine of the  angle between these patterns is $ \mathbf{p}_1\cdot \mathbf{p}_2/(\vert  \mathbf{p}_1\vert \vert \mathbf{p}_2\vert)^{1/2}=W/L$ where $W$ is the number of common active entries between the two patterns.
Let us recode now these patterns  in the much higher dimensional space of all non-ordered entry n-uples, with binary coding such that 1 is obtained when all the $n$ entries of an n-uple are active and zero in all the other cases. In other words, the activity threshold is $n$. In this much higher dimensional space, each recoded pattern has $\begin{pmatrix}L \\n\end{pmatrix}$ active entries while the number of common active entries between the recoded $\mathbf{p}_1$ and $\mathbf{p}_2$ is   $\begin{pmatrix}W\\ n\end{pmatrix}$. Therefore, the cosine of the angle between the recoded patterns is equal to $\begin{pmatrix}W \\n\end{pmatrix}/
\begin{pmatrix}L \\n\end{pmatrix} \sim (W/L)^n$ and quickly tends to zero when n increases. The same argument also shows that the coding level of the recoded patterns is much lower than the original one. In other words, recoding in a high dimensional space tends to orthogonalize  and sparsify patterns, which should lead to an easier separation of different patterns. Marr proposed that this is the essential function of
granule cells with the small difference that assuming they need 3 active mossy fiber inputs to fire, they sample about 4 active 3-uples with their 4 MF inputs. From anatomical data, Marr further estimated that through its
1-2 $10^5$ parallel fiber inputs a Purkinje cell is able to sample about 7000 mossy fibers. This led him to argue that 4 MF inputs per granule cell are optimal for the Purkinje cell to be able to sample a significant fraction of recoded 3-uples.
Albus  noted, based on Cover's celebrated result for the perceptron \cite{cover65}, recalled below, that the number of patterns that can be linearly classified is proportional to the dimension of the activity space, and that indeed linear separation should be easier in the granule cell space than in the mossy fiber ones.

In machine learning, 
projecting the data into a space of large dimension is now a standard algorithmic approach that has been implemented in various ways. For instance, 
Support Vector Machines (SVMs) 
and more generally Kernel methods~\cite{scholkopf01} implicitly realize an embedding of the data in a high dimensional space from which linear separation is more likely to be possible.  Modern artificial networks such as transformers \cite{vaswani17} manipulate vector representations of data in high dimensional spaces. One should note that, in these high dimensional spaces, the relevant information about stimuli/data, the neural dynamics or neural representation associated to a given task,  still live in an underlying space of smaller dimension (independent of the embedding space). Actually, various machine learning techniques, such as auto-encoders, aim at projecting the activity  on a low dimensional space. In 
neuroscience, 
low dimensional dynamics embedded in a high dimensional activity space have been observed in multiple experimental contexts \cite{churchland12,gao15,chung21}, as noted above. These results have triggered multiple studies showing how various types of computations can be performed by networks with low-rank connectivity, which in turn lead to low dimensional dynamics 
(see e.g. Refs. \cite{mastrogiuseppe2018linking,sussillo2013opening}).

Expansion of representation in a higher-dimensional space also appears to be  repeatedly used in neural systems. Beyond the mossy fibers/granule cell relay,  well-known examples include the entorhinal cortex/dentate gyrus 
relay in hippocampus and  the antennal lobe/mushroom body relay  in the fly olfactory system \cite{cayco19}. This has led different studies, both theoretical and experimental, to reexamine the benefits of expansion recoding and sparseness and their realization by cerebellar granule cells. On the theoretical side, Billings et al \cite{Billings2014} found that feed-forward networks with few synaptic connections per output neuron and network-activity dependent threshold were optimal for lossless sparse encoding over a wide range of input activities.  Babadi and Sompolinsky \cite{babadi14} noted that the variability of the inputs should also be taken into account. While  a high-dimensional representation was found to be beneficial for classification, sparseness of this representation was found to amplify the size of fluctuations about the mean input.  For a random linear  projection into a high-dimensional space, they found that a sparseness $f$ of the order of the inverse square root of the dimension $N$ of the high dimensional representation, $f\sim 1\sqrt{N}$, is optimal.
Litwin-Kumar {\em et al}  \cite{litwin17}  reconsidered the determination of the optimal degree of synaptic connectivity for granule cells, and found  that 4-5 inputs ensure that each granule cell neuron will receive a unique set of inputs, for realistic values of the numbers of  mossy fiber inputs and granule cells, and that the dimensionality of the granule cell representation of the inputs is  close to being maximized.

To assess the impact of correlation of different granule cell activities,  Cayco Gajic {\em et al} \cite{cayco17} performed simulations of a flat layer of granule cells with randomly located mossy fiber rosettes with realistic density and realistic granule cell dendrites. Network performance is assessed by the learning speed of pattern separation of a back-propagation algorithm (admittedly a non-biological feature). Sparseness of activity in the granule cell layer is found to increase learning speed and reduce correlation of activity at the population level. On the experimental side, progress in imaging neural activity has led different groups to attempt and measure the sparseness and the dimension of  the granule cell population activity. Using  calcium indicators and two-photon microscopy,
different groups have reported both in mice \cite{giovannucci17,wagner17} and in  larval zebrafish \cite{knogler17} that the activity in the granule cell population was dense and low-dimensional. However, a more recent study~\cite{lanore2021} found  sparse high dimensional representions  in imaging parallel fiber activity in mice, with one dimension for 5 parallel fiber suggested to explain the variance of population activity. The difference with previous results are attributed by the authors of ref.~\cite{lanore2021} to the intrinsic low-dimensionality of behavioral tasks in previous experiments, while their mice were headfixed but free to run on a wheel or whisk. Another recent study has also demonstrated sparse activity of granule cells, and showed in addition that sparseness is ensured by inhibition from Golgi cells \cite{fleming24}.

\section{The Purkinje cell as a Perceptron.}

\begin{figure}
\begin{tabular}{ll}
{\bf A} & {\bf B} \\
\begin{minipage}{0.64\textwidth}
\includegraphics[width=\textwidth]{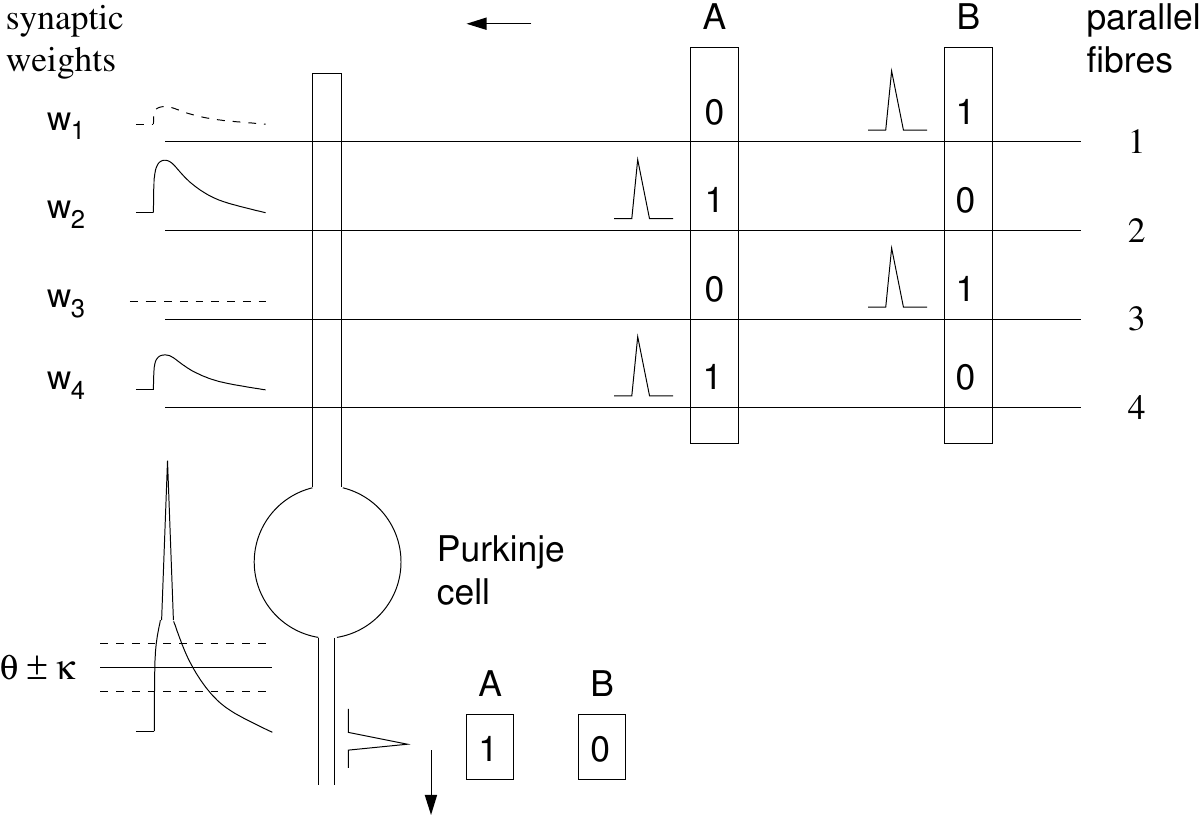} 
\end{minipage} & 
\begin{minipage}{0.35\textwidth}
\includegraphics[width=\textwidth]{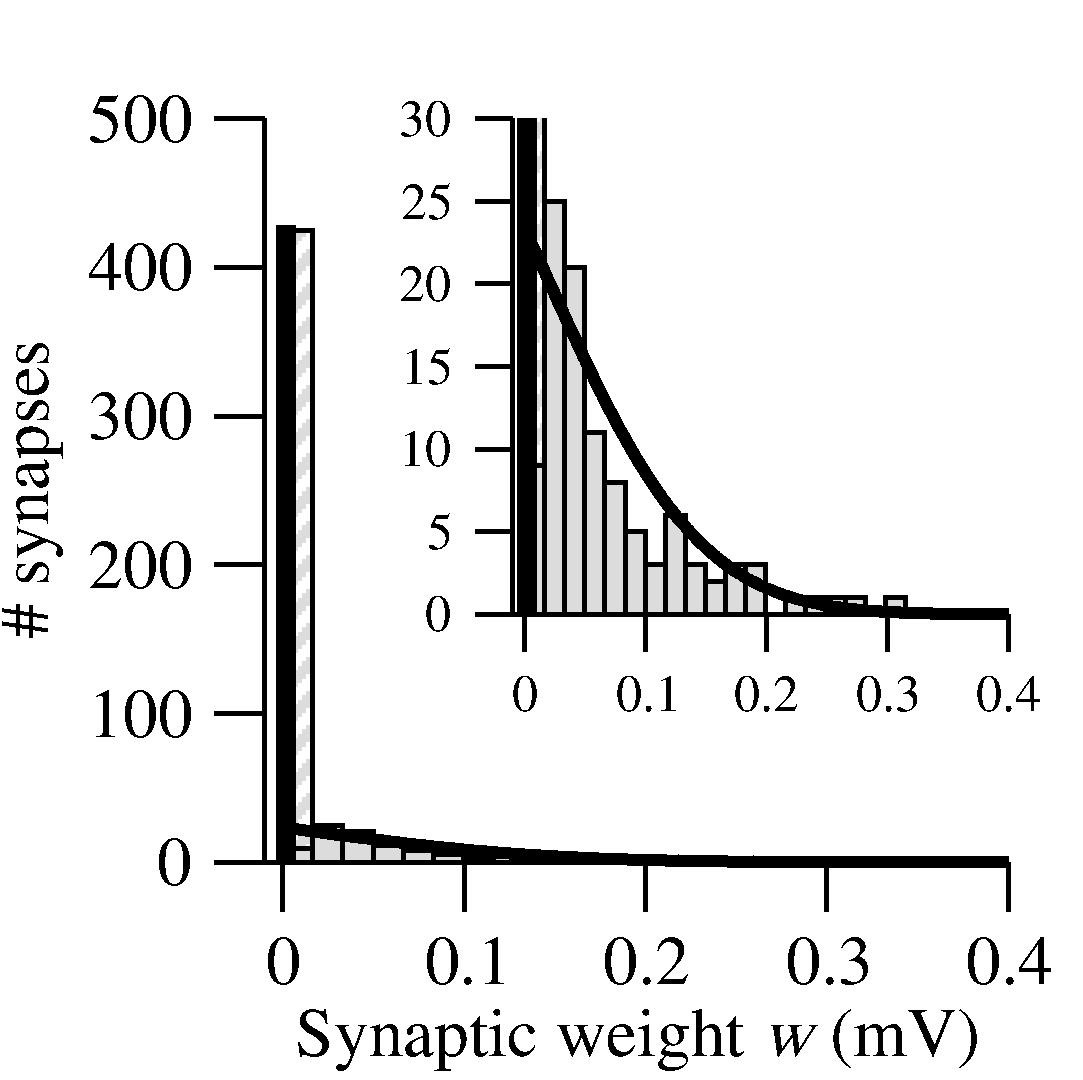}
\end{minipage}
\end{tabular}
\caption[]{The Purkinje cell as a Perceptron. A: Sketch of a perceptron with granule cell inputs, tasked with learning input-output associations (here, 4 inputs and 2 associations are shown). B: Distribution of synaptic weights at maximal capacity: Experimental data (measurements of synaptic weights using recordings of granule cell-Purkinje cell pairs) vs Analytical distribution of weights for a perceptron with sign-constrained weights, with parameters that best fit the data. Figure redrawn from \cite{brunel04}.}
\label{fig:PC-perceptron}
\end{figure}

Marr and Albus proposed that learning by a Purkinje cell could operate similarly as a  perceptron (Fig. \ref{fig:PC-perceptron}A).
In 1958 Franck Rosenblatt~\cite{rosenblatt58,rosenblatt62}, a psychologist, introduced the perceptron as a simple neural model with the initial goal of understanding how the brain learns a task from examples. Actually, Rosenblatt considered a large family of possible neural architectures, but the simplest, purely feed-forward, one is what kept the name of perceptron. This perceptron has a number $N$ of inputs cells, directly feeding a layer of output neural cells. These cells have binary activities (resting state/active state), the active state being produced if the linear combination of the inputs, each weighted by a synaptic efficacy, is greater than a threshold. Since output units only receive information from the input layer (no lateral connections), learning operates independently in different output neurons, and thus one can consider a single output unit. For such unit, Rosenblatt proposed a supervised learning algorithm, the perceptron algorithm, which allows the unit to find synaptic weights for which a given set of (input pattern, output activity) associations are realized - provided that a solution exists. The capacity of the perceptron is the maximum number $p_{\text{max}}$ of associations that the perceptron can learn, as a function of its size $N$. This capacity has been studied in the 60's with a geometrical approach by Cover~\cite{cover65} for input patterns 'in general position' (no linear dependencies), within the context of statistical learning theory~\cite{vapnik99} ({\em Vapnik-Chervonenkis dimension}), and for specific Hebbian learning rules notably by Willshaw {\em et al}~\cite{willshaw69}. In the late 80's, Elizabeth Gardner~\cite{gardner88,gardnerderrida88} showed how modern tools from the statistical physics of disordered systems can be used to compute the capacity in the limit of very large network size, for randomly chosen patterns and associations. In this large size limit, the capacity scales with network size, and one computes the {\em critical capacity} $\alpha_c=\lim_{N\rightarrow \infty} p_{\text{max}}/N$. 
 Making use of these statistical physics methods, Brunel {\em et al}~\cite{brunel04,barbour07} have reconsidered the analysis of the perceptron as a model for the Purkinje cell. The large size limit is relevant here since the number of inputs (paralell fibers) to a PC is of order $150, 000$. Moreover, the method introduced by Elizabeth Gardner is able to take into account various constraints. In particular, as discussed above for the PC, standard models assume that the synapses subject to weight modifications during learning are excitatory. Hence in the model the synaptic weights must be kept non-negative.  Statistical physics methods also makes it possible to compute the synaptic weight distribution after learning. 

 These calculations can be compared with experimental recordings of synaptic strengths of parallel fiber to Purkinje cell synapses,
 performed by Isope and Barbour~\cite{isope02b} in adult rat cerebellar slices (Fig. \ref{fig:PC-perceptron}B), using paired recordings of granule cell (in loose cell-attached mode) and Purkinje cells (in whole-cell patch-clamp mode).  To confront the model with these experimental data, and infer unknown parameter values, the  synaptic weights in the model are assumed to correspond to the peak somatic depolarization triggered by a granule cell action potential. The experiments provide the distribution of these weights and a somewhat unexpected result. From earlier anatomical studies \cite{harvey88}, one expects $\sim 50\%$ of `on beam' parallel fibers from presynaptic granule cells to make actual synaptic contact with the PC. However, electrophysiological recordings lead to noticeable synaptic currents in only $\sim 10\%$ of pairs, instead of the expected  $\sim 50\%$. This implies that $80\%$  of the synapses are `silent' (i.e. anatomically present, but leading to no detectable  response). 
 In the model, the  capacity is computed under the constraint of positive synaptic weights, in the large size limit. In such limit, one expects that at  maximum capacity a large fraction of the constraints are saturated, hence a large fraction of synapses should be silent. The theoretical computation indeed gives that the distribution of weights at maximum capacity has two components, a fraction of $50\%$ of silent synapses, and a truncated Gaussian distribution for the non-zero weights.  Adding additional constraints, such as a level of robustness to noise,  increases the fraction of silent synapses (Fig. \ref{fig:PC-perceptron}B) . The computation can also be performed below maximum capacity. The distribution in this case has peak near zero, that becomes sharper as one gets closer to the maximum capacity. Fitting the model to the empirical data, one gets estimates of the parameter characterizing the robustness to noise, of the fraction of active inputs and of active outputs. 
 The perceptron learning algorithm provides a supervised scheme for learning input-output associations, in the absence of any additional constraint. 
 This learning algorithm can be adapted to the case of sign constrained synaptic weights \cite{brunel04,clopath12}.
This suggests that a biologically plausible learning scheme could allow Purkinje cells to reach their maximum  capacity with positive synaptic weights. 

In the above approach, one unrealistic assumption is that learned associations are statistically independent. Clopath {\em et al} \cite{clopath12} have considered an extension of the model in which sequences of patterns are learned by the perceptron, with statistical correlations between successive patterns and between successive desired outputs. They show that capacity increases with the level of both input and output correlations. They also considered a scenario in which PCs are bistable, and showed that the capacity is increased if the correlation in  the output is larger than the correlation in the input. Bi-stability means here that there is a gap between the input current strength needed to go from the down to the up state, and the one needed to go from the up to the down state. There is thus a range of currents values for which the cell is found in either the up or down state depending on past history.
Clopath and Brunel~\cite{clopath13} have studied the capacity and synapse distribution of an analog perceptron with excitatory synapses. The qualitative results are similar to the ones for the binary perceptron, with an additional feature that the maximum capacity is optimized with a sparse input distribution.

Learning capabilities of Purkinje cells have also been studied in much more biophysically detailed models. Steuber et al \cite{Steuber2007} trained a detailed Purkinje cell model with specific granule cell input patterns, using LTD triggered by coincident granule cell and climbing fiber input. They then found that learned inputs could be distinguished from other non-learned inputs using the duration of a pause following a burst of spikes triggered by the pattern. Learned patterns elicited shorter pauses, hence a larger Purkinje cell output. Some of the predictions of this model were tested experimentally both in vitro and in vivo. Safaryan et al \cite{Safaryan2017} studied instead the effect of non-specific  plasticity on a similar detailed Purkinje cell model, and showed that it improves the recognition of sparse patterns degraded by local noise. There is currently a large gap between studies of learning in detailed biophysical models and in highly simplified binary or analog models, and a central challenge for theorists in future years will be to bridge this gap.

\section{Temporal basis functions.}

\begin{figure}
\includegraphics[width=\textwidth]{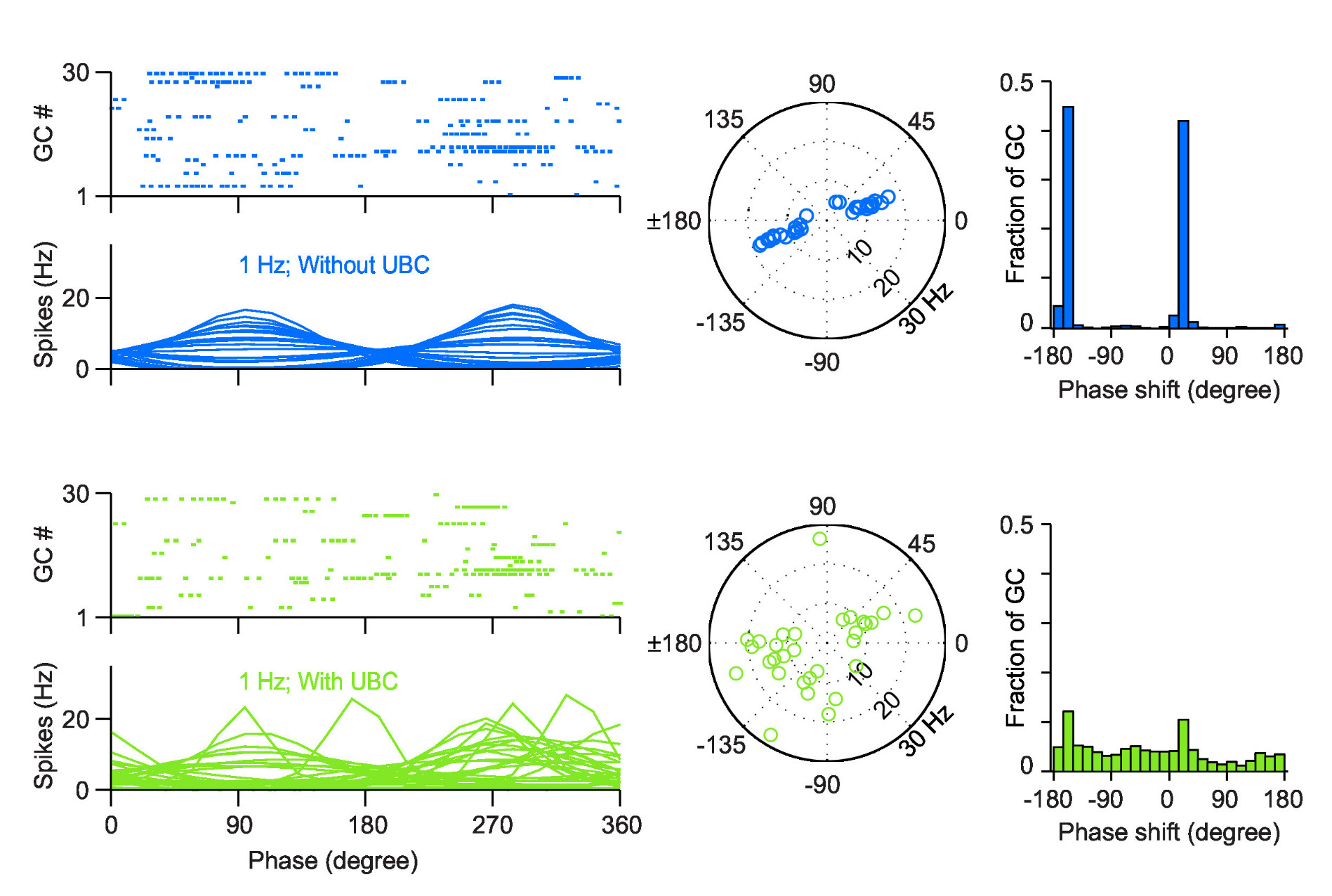}
\caption{Simulation of a Granular network with or without UBCs, that are characterized by a broad diversity of synaptic time constants. Top: In the absence of UBCs, the granule cell network responds in a highly stereotyped fashion to a 1Hz sinusoidal input coming from MFs (top left: raster plot; bottom left: average spike rates of different cells in the network; center: Polar plot of the response of granule cells to sinusoidal inputs; right: Distribution of phase shifts of granule cell firing with respect to the inputs). Bottom: In the presence of UBCs, granule cells respond in a highly heterogeneous way, with a broad diversity of phase shifts. This heterogeneity can then be exploited by Purkinje cells to learn input/output relationships with arbitrary temporal structure, by selecting appropriate subsets of granule cells through supervised synaptic plasticity. From \cite{zampini16}.}
\label{fig:time-ubc}
\end{figure}

The models of Marr and Albus describe the cerebellum as a pattern associator, tasked with learning the correct sensory input to motor output associations. A major drawback of these models is that the Purkinje cell response depends only on the currently available  mossy fiber input. 
However, in many motor behaviors, the adequate motor response depends on sensory inputs that are no longer available. In the paradigmatic eyeblink conditioning example from the cerebellar literature mentioned above, animals need to learn the correct timing between a sensory stimulus and an airpuff applied to the eye so that they can blink at the appropriate time to avoid the airpuff. Although mossy fiber inputs are expected to signal the sensory stimulus, this signal is no longer expected to be present at the time of the airpuff. Thus, there is a need to maintain sensory information on a range of time scales so that a correctly timed motor output can be generated.

This issue motivated Fujita \cite{fujita82} to propose his ``adaptive filter model'' of the cerebellum, where sensory inputs are filtered using a diversity of time constants, so that Purkinje cells can learn the correct set of filters for a desired movement. The idea is that somewhere in the cerebellum in between mossy fibers and Purkinje cells, a set of temporal basis functions are generated \cite{buonomano94,Medina2000,porrill07}.

What could be the mechanisms for generating a diversity of time scales that can be exploited by downstream Purkinje cells? Multiple potential loci have been proposed and characterized experimentally in recent years. First, mossy fibers themselves could have an extended range of time scales. Second, UBCs, which receive mossy fiber inputs and project in turn to granule cells, have been shown to be equipped with a strong diversity of glutamatergic receptors (AMPA, NMDA, mGluR1, mGluR2) that are expressed in a highly heterogeneous fashion, leading to broad diversity of temporal responses to MF inputs (Fig.~\ref{fig:time-ubc}) in both electric fish \cite{kennedy14} and rodents \cite{zampini16,guo21}. Third, short-term plasticity at the MF-GC synapse has been shown to exhibit a broad diversity of time scales \cite{chabrol15,barri22}. Fourth, as initially proposed by Fujita \cite{fujita82}, recurrent inhibition from Golgi cells could in principle lead to temporally complex dynamics that could also allow the granule cell layer network to maintain sensory information on multiple time scales (see e.g.~\cite{dean10,rossert15}). Fifth, long time scales could be present in Purkinje cells themselves, using for instance metabotropic glutamate receptors \cite{Fiala1996,Steuber2004}. Thus, multiple potential mechanisms could allow Purkinje cells to learn input/output associations that require integration of inputs over multiple time scales.

\section{Learning : complex spikes and plasticity of granule cell to Purkinje cell synapses}

\begin{figure}
\begin{tabular}{ll}
{\bf A} & {\bf B} \\
\begin{minipage}{0.28\textwidth}
\includegraphics[width=\textwidth]{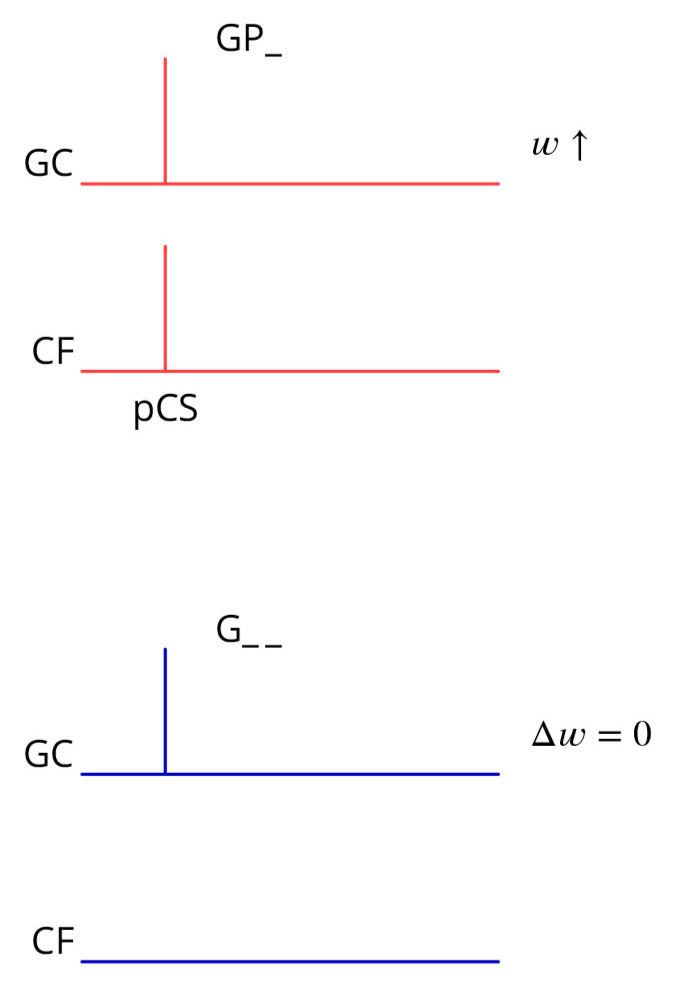} 
\includegraphics[width=\textwidth]{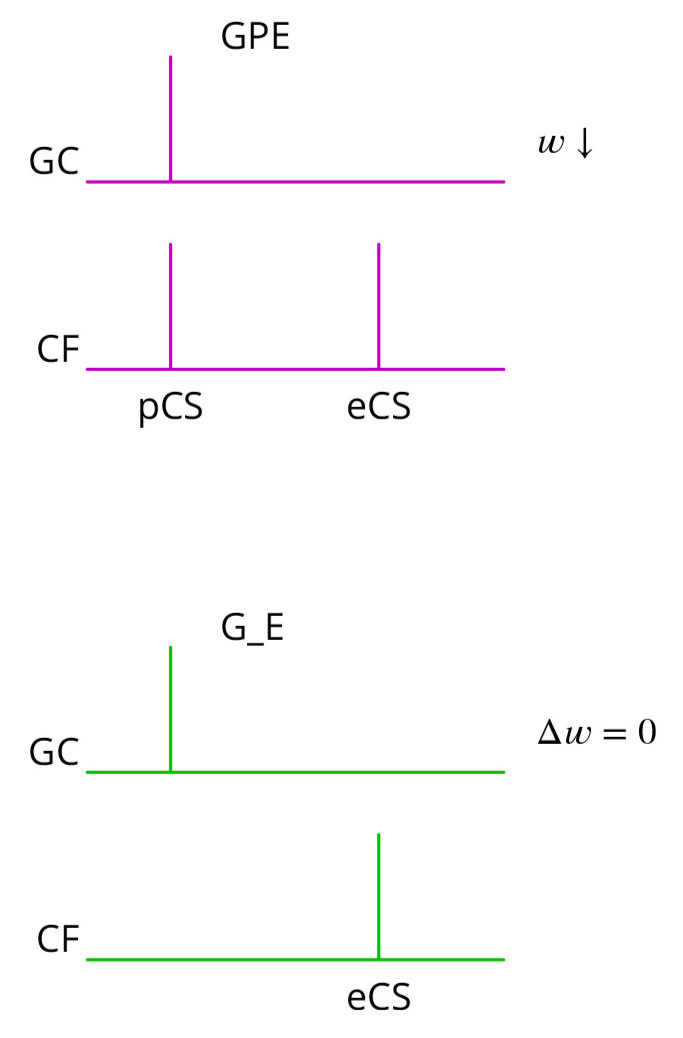} 
\end{minipage} & 
\begin{minipage}{0.7\textwidth}
\includegraphics[width=\textwidth]{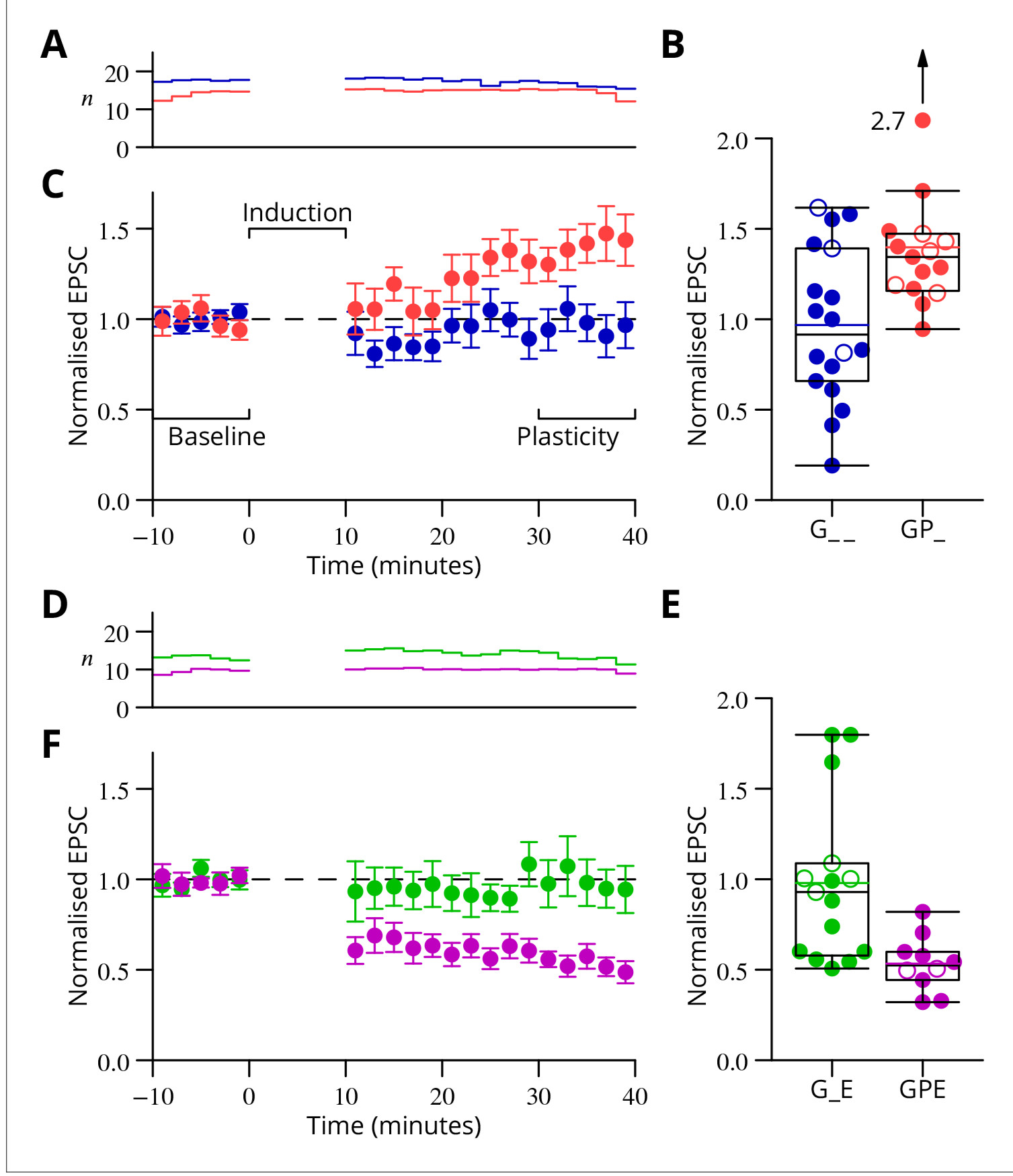}
\end{minipage}
\end{tabular}
\caption[]{A. Plasticity rule predicted by a stochastic gradient descent algorithm using climbing fibers as both the perturbation and the error signal (pCS: perturbation Complex Spike; eCS: error Complex Spike). B. Plasticity rule observed in vitro in physiological calcium. All four possible patterns shown in A produce the predicted outcomes (same color codes as in A). From \cite{bouvier18}.}
\label{fig:plasticity}
\end{figure}

Two of the major predictions of Marr-Albus theories is that climbing fibers (CFs) should provide a `teaching' or `error' signal to Purkinje cells, and that this signal should modify the synapses of active Granule cell inputs. On the one hand, the prediction that CFs control plasticity at PF$\rightarrow$ PC synapses was confirmed experimentally by Ito, in the early 1980s, in {\em in vitro} cerebellar preparations \cite{ito82}, and by a multitude of other groups since. It has also been shown that the presence of CFs on a given trial influences PC firing on the next trial in vivo, consistent with plasticity induced by the CF signal \cite{yang14}. On the other hand, the question whether CFs provide an error or teaching signal has been more difficult to answer. In some experimental conditions, CFs do provide error information \cite{herzfeld18}.
However, it has also been shown that CFs can carry information about other variables, such as different  reward-related variables (expectation, delivery,...) or behavioral variables (see e.g.~\cite{hull20,Kostadinov2019,Ikezoe2023}).

On the theoretical side, an outstanding question is how the cerebellum solves the `credit assignment' problem. For supervised learning to be effective, each Purkinje cell should receive its own dedicated error signal, that should convey in particular the sign of synaptic changes that lead to a decrease in motor error. Cerebellar learning has mostly been considered experimentally in the context of simple and stereotyped movements where error signals are easy to compute and to signal to relevant Purkjinje cells. Examples of such simple movements are eye movements, where `retinal slip' signals can be easily computed and fed back to the cerebellum to learn to adapt the vestibulo-ocular reflex \cite{robinson76,ito74,blazquez04}, or to adjust the location of saccades \cite{optican80,dash14,soetedjo08}. In more complex movements, in particular those that require different Purkinje cells to adjust in different ways, it is unclear how error signals that are specific for each cell can be computed.

Recently, Bouvier {\em et al}  \cite{bouvier18} have proposed a learning algorithm that can solve this credit assignment problem. This algorithm belongs to the general class of stochastic gradient descent algorithms \cite{williams92,xie04}. The idea is to have a source of stochastic perturbations of movements, and to reinforce perturbations that improve movement, while penalizing perturbations that lead to worse performance. They proposed that climbing fibers could also act as the stochastic perturbations, with the presence of absence of a second complex spike signaling whether the perturbation was bad (thus needing depression of active inputs) or good (thus needing potentiation of active inputs), respectively. In this view, climber fiber activity represents a combination of a source of noise, and of an error feedback. Bouvier {\em et al} \cite{bouvier18} showed that the resulting plasticity rule is consistent with in vitro plasticity experiments (Fig.\ref{fig:plasticity}), and analyzed the convergence and capacity of this algorithm. While theoretically appealing, it remains to be demonstrated that this algorithm operates in vivo in the cerebellum during learning of any motor behavior. 

Although most of the field's attention has focused on the granule cell to Purkinke cell synapse as the main site of learning, multiple experiments have demonstrated that plasticity must occur at other sites. In particular, Miles and Lisberger proposed that while learning occurs initially at the GC-PC synapse, it is then transferred to the deep cerebellar nuclei. This process has been studied in multiple theoretical models (e.g.~ \cite{clopath14}).
Finally, we note that many other types of synapses have been shown to be plastic (see e.g.~\cite{hansel01,hull22,Gao2012}), 
and it remains to be understood if and how these other plasticity sites contribute to learning. 

\section{Conclusions and perspectives}

Much more information is available today on neural systems than 40 years ago, when G\'erard Toulouse and a few other physicists turned their attention to neuroscience, stimulated by Hopfield's pioneering work on associative memory models \cite{hopfield82}.  Similarly to other brain areas, progress in experimental techniques has since made available a wealth of data on the cerebellum including detailed descriptions of different cell types \cite{kozareva20}, 
connectomic reconstructions \cite{nguyen23}, 
and simultaneous recordings of large numbers of neurons in awake behaving animals \cite{beau24}.  
Modern tracing techniques also allow one to monitor and stimulate the neural areas targeting the cerebellum as well as its DCN outputs. On the theoretical side, old ideas have been clarified and novel ones have been proposed, as summarized in part in the present short survey. It is however clear that much remains to be done to obtain a full understanding of the role of the cerebellum 
in how the brain process information, learns and controls behavior, be it motor or otherwise.

One main question is whether and how the regular architecture of the cerebellum
is used beyond the simple stereotypic movements that have mainly be studied. How is the canonical cerebellar circuit used and repurposed for  tasks ranging from complex movements, to language processing up to social interactions? Proposals range from the wide utility of fine discrimination through high dimensional expansion, to the building of predictive internal models \cite{ito08,hull20} and the ability of the cerebellar circuitry to perform universal function approximation \cite{caycogajic24}. 
A complete account of cerebellar function should certainly include
its dynamics. This includes the question of the functional importance of the different cerebellar rhythms which have been recorded for a long time \cite{adrian35}, the basis of which has been in part elucidated \cite{desolages08,dugue09} but the role of which is still unclear, as in many other neural structures. 

Learning is another important question that is still poorly understood. 
There is no evidence that the brain makes use of the same type of gradient descent learning algorithms as those so successfully applied in machine learning. And if so, it is generally unclear how such algorithms can be implemented in biological neural networks \cite{lillicrap20}.
The cerebellum hopefully offers a particularly promising example  to go beyond Marr-Albus-Ito theory and decipher how credit-assignment is performed biologically.
At a larger scale, it remains to  better understand how different neural areas coordinate their activity to perform a definite function. Movement performance and its sharing between (at least) the motor cortex, the basal ganglia and the cerebellum appears a particularly important case to elucidate, both for physiology and pathology. 
We certainly hope that joint experimental and theoretical efforts will overcome these challenges in the coming decades, finally fulfilling G\'erard's dream.

\begin{acknowledgments}

We are grateful to our friends and colleagues, Yonatan Aljadeff, Philippe Ascher, Boris Barbour, Guy Bouvier, Mariano Casado, Claudia Clopath, Chris De Zeeuw, Stéphane Dieudonné, David Di Gregorio, Guillaume Dugu\'e, Christian Hansel, David Herzfeld, David Higgins, Court Hull, Clément Léna, Stephen Lisberger, Alain Marty, Wade Regehr, and Vincent Villette for many insightful discussions over the last 25 years.

\end{acknowledgments}


\end{document}